\documentclass[fleqn,usenatbib]{mnras}

\usepackage{newtxtext,newtxmath}
\usepackage[normalem]{ulem}
\usepackage{kantlipsum}
 \usepackage{amsmath,array}

\usepackage[T1]{fontenc}

 \usepackage{xcolor}
\usepackage{graphicx}	
\usepackage{amsmath}	

\newcommand{\rsun}{\ensuremath{\mathit{R}_{\odot}}}                  

\newcommand{\mdot}{\ensuremath{\dot{M}}}                             
\newcommand{\rstar}{\ensuremath{\mathit{R}_{\star}}}                 
\newcommand{\vinf}{\ensuremath{v_{\infty}}}                          %

\title[The disappearing massive LBV star in PHL~293B]{The possible disappearance of a massive star in the low metallicity galaxy PHL~293B\thanks{Based on observations collected at the European Southern Observatory under programmes 70.B-0717, 60.A-9442, 60.A-9502, 2104.D-5015.}}

\author[A. P. Allan et al.]{
Andrew P. Allan,$^{1}$
Jose H. Groh,$^{1}$
Andrea Mehner,$^{2}$
Nathan Smith,$^{3}$
Ioana Boian,$^{1}$
\newauthor Eoin J. Farrell$^{1}$
and Jennifer E. Andrews$^{3}$
\\
$^{1}$School of Physics, Trinity College Dublin, 
   the University of Dublin, College Green, Dublin-2, Ireland; email: allana@tcd.ie\\
$^{2}$ESO -- European Organisation for Astronomical Research in the Southern Hemisphere, Alonso de C\'{o}rdova 3107, Vitacura, Santiago de Chile, Chile\\
$^{3}$Steward Observatory, University of Arizona, 933 N. Cherry Ave., Tucson, AZ 85721, USA
}

\date{Accepted XXX. Received YYY; in original form ZZZ}

\pubyear{2020}

 \hypersetup{draft}
\begin{document}
\label{firstpage}
\pagerange{\pageref{firstpage}--\pageref{lastpage}}
\maketitle

\begin{abstract}
We investigate a suspected very massive star in one of the most metal-poor dwarf galaxies, PHL~293B. Excitingly, we find the sudden disappearance of the stellar signatures from our 2019 spectra, in particular the broad H lines with P~Cygni profiles that have been associated with a massive luminous blue variable (LBV) star. Such features are absent from our spectra obtained in 2019 with the ESPRESSO and X-shooter instruments of the ESO's VLT. We compute radiative transfer models using CMFGEN that fit the observed spectrum of the LBV and are consistent with ground-based and archival Hubble Space Telescope photometry. Our models show that during 2001--2011 the LBV had a luminosity $L_* = 2.5-3.5 \times 10^6 ~L_{\odot}$, a mass-loss rate $\dot{M} = 0.005-0.020 ~M_{\odot}$~yr$^{-1}$, a wind velocity of 1000~km~s$^{-1}$, and  effective and stellar temperatures of  $T_\mathrm{eff} = 6000-6800$~K and $T_\mathrm{*}=9500-15000$~K. These stellar properties indicate an eruptive state. We consider two main hypotheses for the absence of the broad emission components from the spectra obtained since 2011. One possibility is that we are seeing the end of an LBV eruption of a surviving star, with a mild drop in luminosity, a shift to hotter  effective  temperatures,  and  some  dust obscuration. Alternatively, the LBV could have collapsed to a massive black hole without the production of a bright supernova.
\end{abstract}



\section{Introduction}

Massive stars are among the most important sources of ionising photons and chemical elements, producing a significant proportion of the elements currently present in the Universe. They are instrumental to the understanding of a variety of astrophysical topics, including the link between supernovae (SNe) and gamma-ray bursts to the nature of their respective progenitors \citep[e.g.,][]{Schulze_2015}), as well as the early evolution of the Universe. 

Our current understanding of massive stars and their fates is quite incomplete in environments with metallicity ($Z$) lower than the Small Magellanic Cloud \citep[$Z\simeq0.2Z_{\odot}$][]{Hunter_2007}. This is owed mainly to a scarcity of observations of massive stars at very low $Z$, especially in late evolutionary stages. Wolf-Rayet (WR) stars are generally rare in very metal-poor regions \citep{Crowther_Hadfield_2006}, with surveys detecting evolved WR stars in the SMC \citep{Massey, Massey_2014, Neugent_2018, 2020_Shenar} and  I~Zw~18 galaxy \citep{Izotov_1997,  Legrand_1997, Brown_2002}. A handful of luminous blue variables have been found in the SMC including HD5980 \citep{Drissen_2001, Barba_1995} and other low-metallicity galaxies \citep{Izotov_2009, X}. Red supergiants are also thought to be rare at very low $Z$ \citep[e.g.][]{Eldridge_2017}.

Efforts to advance our understanding are being made, with recent numerical stellar evolution models of low-metallicity stars revealing a surprising prediction. They indicate that some of the most massive stars may end their lives as unstable LBV stars \citep{Groh_GRID_Z_0.0004}, as they fail to shed mass and become H-poor WR stars. The LBV phase is thought to occur late in the evolution of massive stars \citep[e.g.,][]{Humphrey_davidson, Maeder_Meynet, Groh_2014_60}. LBVs show re-occurring eruptive events generating considerable mass loss \citep{smith_owocki2006}, in addition to irregular photometric and spectroscopic variations of the S-Doradus type \citep{vanGenderen2001}. LBVs play a key role in the mass budget of very massive stars \citep{Smith2014ARAA} and regulate their final compact remnant masses, which in some cases can be a massive black hole \citep{2019arXiv191200994G}. LBVs are also thought to be immediate progenitors of some SN explosions \citep[e.g.][]{kotak_Vink_2006, Gal_yam_2009_2005gl,Smith_2011,Groh2013a, boian_2018_catch}.

In an effort to improve our understanding of very massive stars at low $Z$, we have monitored the blue compact dwarf (BCD) galaxy PHL~293B. This galaxy lies at a distance of 23.1 Mpc \citep{Mould}, and has a metallicity of $Z\simeq0.1Z_{\odot}$ \citep{Izo_thuan_st_2007}. Spectroscopic observations of the compact galaxy obtained between 2001--2011 consistently featured broad, strong emission components in the hydrogen Balmer lines. These spectral features have been interpreted to originate in the LBV outflow \citep{Izotov_2009,X}, since together with the presence of \ion{Fe}{II} and weak \ion{He}{I} lines, only LBVs show these types of signatures \citep[see discussion in Sects. 2.4 and 3 of][]{Groh_2014_60}. These earlier spectra were remarkably similar, differing mainly in the strength of the narrow components, likely due to the different aperture sizes used. Photometric analysis of PHL~293B revealed no optical photometric variability at the level of 0.1 mag between 1988 and 2013 \citep{Terlevich_shell}. Based on this, \citet{Terlevich_shell} suggested that the blueshifted absorptions of H~I and Fe~II were not caused by an LBV, but instead by an expanding supershell generated by the cluster wind of PHL~293B. The substantial spectral variation we report disfavours such a hypothesis. \citet{Burke2020} also report the weakening of the hydrogen broad components based on 2019 Gemini data. In addition, they report photometric variability of 0.12~mag in $g$ band between 1998 and 2018, using images from the Sloan Digital Sky Survey (SDSS) and the Dark Energy Survey (DES). They suggest a SN~IIn or an unusual outburst as the source of the broad components in the 2001--2011 spectra.

\begin{figure}
\includegraphics[scale=0.51]{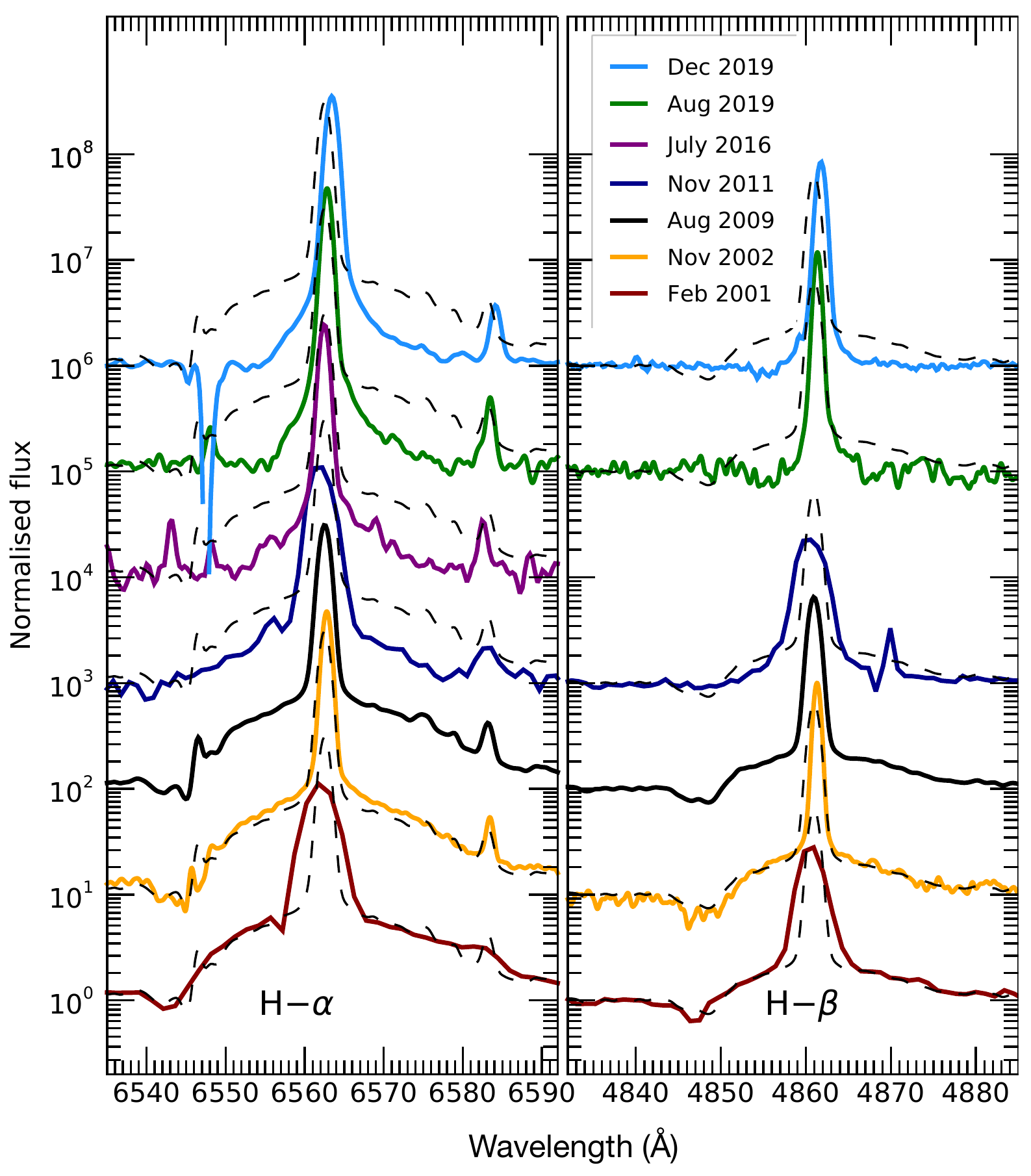}
\caption{Spectroscopic evolution of PHL~293B between 2001 (lowest spectrum) and 2019 (highest). The 2009 X-shooter spectrum (dashed grey) is overplotted to all spectra, which are also shifted for clarity. }
\label{fig: obs}
\end{figure}  

Here, we report new observations of PHL~293B obtained in 2019 with the European Southern Observatory's Very Large Telescope (ESO/VLT) instruments X-shooter and Echelle Spectrograph for Rocky Exoplanet- and Stable Spectroscopic Observation (ESPRESSO). We also discuss unpublished, optical archival imaging obtained with the {\it Hubble Space Telescope (HST)}. We compute radiative transfer models of stellar winds to interpret the spectroscopic observations and the existing photometric data. As we elaborate below, we find that the LBV was in an eruptive state at least between 2001--2011, which then ended, and may have been followed by a collapse into a massive black hole (BH) without the production of a SN. This scenario is consistent with the available {\it HST} and ground-based photometry.

\section{Observations and modelling of the LBV} \label{Section_obs_mod}

\begin{figure}
\includegraphics[scale=0.54]{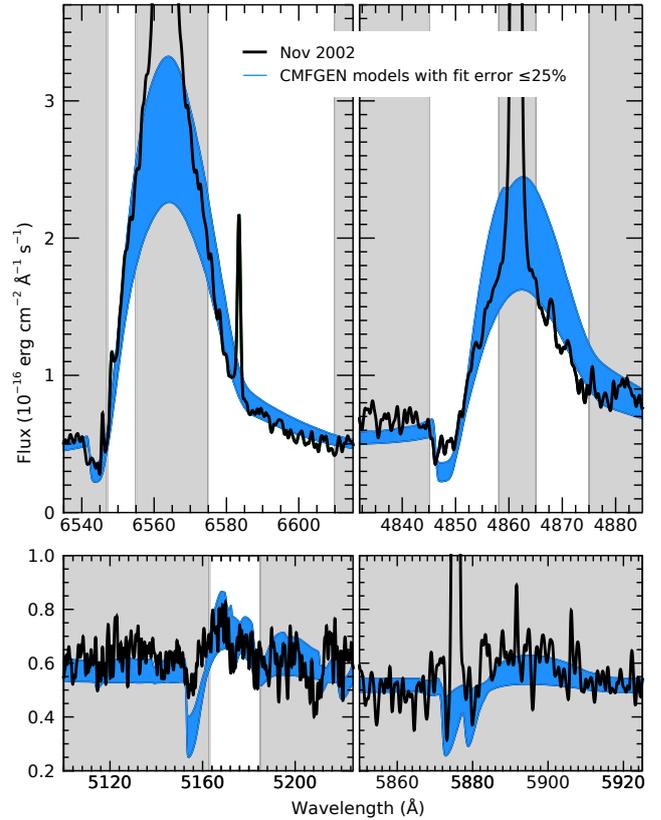}
\caption{Comparison of the 2002 UVES spectrum of PHL~293B (black) to our best-fitting CMFGEN models. Blue region corresponds to models with a fit error $\leq 25\%$. White regions show the wavelength ranges considered in the fit. The upper left and right panels display the H-$\alpha$ and H-$\beta$ emission lines respectively.  The lower left panel includes the tested \ion{Fe}{ii} 5169~\AA\ emission line in addition to weaker \ion{Fe}{ii} 5159~\AA\ and \ion{N}{i} 5198~\AA\ emission and \ion{Fe}{iii} 5156~\AA\ absorption. Some best fit models also feature weak \ion{Mg}{i} 5173~\AA\ and 5184~\AA\ and \ion{Tk}{ii} 5186~\AA\ and 5189~\AA. The lower right panel includes the \ion{He}{i} 5786~\AA\ line as well as \ion{Na}{i} 5890~\AA\ and 5896~\AA\ emission.}
\label{fig: model}
\end{figure}

\begin{table*}
	\centering
	\caption{Summary of the spectroscopic observations of PHL~293B from 2001 to 2019.}
	\label{tab:tab_obs}
	\begin{tabular}{lccccc} 
		\hline
		Instrument  & Date & Wavelength range (\AA)  &  Setup & Resolving power \\
		\hline 
			 VLT/X-shooter &  2019-12-17 &  3000--23000  & UVB, VIS, NIR (1.0'', 0.9'', 0.9'') & 5400, 8900, 5600  \\
			 VLT/ESPRESSO &  2019-08-28 &  3800--7800   & MR42 (4x1'') & 72\,000    \\  
				HST/COS &  2018-04-30 &  1027--2337  & G140L (2.5'')  &  2000\\
				INT/IDS &  2016-07-02 &  5890--7230   & H1800V (1'') & 7000 \\
                 WHT/ISIS &  2011-11-30 &  3805--8095   & R300B/7500 (1'') & 1000
                    \\
                      WHT/ISIS &  2011-09-24 &  2630--6230  & R300B/5700 (1'') & 1000
                    \\
              VLT/X-shooter &  2009-09-28 &  3000--23000   & UVB, VIS, NIR (1.0'', 0.9'', 0.9'')& 5100, 8800, 5600   \\
 			  VLT/X-shooter &  2009-08-16 &  3000--23000   & UVB, VIS, NIR (1.0'', 0.9'', 0.9'') & 5100, 8800, 5600  \\
 			  VLT/UVES &  2002-11-08   & 3100--6800  & DICHR\#1 (1'') &  40\,000  \\
 			  SDSS &  2001-02-22 &  3810--9200  & 3'' fiber & 2000 \\
		\hline
			\end{tabular}
\end{table*}

We present new spectra of PHL~293B obtained in 2019 using the ESO/VLT instruments X-shooter and ESPRESSO and compare these to archival spectra obtained between 2001 and 2016 (see Table \ref{tab:tab_obs}). ESPRESSO was used in the four-Unit Telescope mode, combining the light from four 8~m telescopes. The data were reduced with the ESO pipeline version espdr/1.3.0. The X-shooter data were reduced with the ESO pipeline version xshoo/3.3.5. The archival INT/IDS, WHT/ISIS, and 2009 X-shooter spectra were obtained without flux standards and flat fields. Therefore, we re-scaled the flux as to best match the equivalent width of the narrow component of the H-$\alpha$ line of a spectrum with a similar aperture size (UVES). 

Figure \ref{fig: obs} highlights the absence of the inferred LBV signature from the spectra of PHL~293B in 2016 to 2019. Other signatures that were interpreted as being due to the LBV, such as from \ion{Fe}{ii} and \ion{He}{i} \citep{Izotov_2009,X}, are also not detected in our 2019 data. We interpret the lack of spectral variability in the 2001, 2002, 2009 and 2011 spectra \citep[also noted by][]{Terlevich_shell} as confirmation of the LBV's presence, with the inferred LBV signatures disappearing sometime between 2011 and 2016.

We also obtained archival {\it HST} imaging of the PHL~293B galaxy, which was observed on 2010 October 31 with {\it HST}/WFC3 with filters F336W, F438W, F606W, and F814W (GO-12018; PI Prestwich). We performed photometry of the brightest region of the galaxy, where the LBV is thought to reside since this is where previous spectroscopic observations centred their slit on. We use Dolphot \citep{Dolphin2000,Dolphin2016} on the standard STScI pipeline pre-processed, charge-transfer efficiency corrected images. These observations were obtained before the inferred LBV signature within the spectra disappeared, and obtain $m_{\text{F336W,pre}}=19.70 \pm 0.01$, $ m_{\text{F438W,pre}}=20.82 \pm 0.01$, $ m_{\text{F606W,pre}}=20.31 \pm 0.01$, and $m_{\text{F814W,pre}}=20.03 \pm 0.01 $~mag in the F336W, F438W, F606W, and F814W filters, respectively.

To constrain the parameters of the LBV in PHL~293B during outburst, we compute new radiative transfer models and fit the high-resolution 2002 UVES spectrum. We use the line-blanketed atmospheric/wind radiative transfer code CMFGEN \citep{hill_mill} to compute continuum and line formation in non-local thermodynamic equilibrium (NLTE) and spherical symmetry. Our new models use similar physical assumptions to those of the LBV progenitor of the SN candidate SN~2015bh \citep{boian_2018_catch}. CMFGEN takes as input the stellar luminosity $L_*$,  stellar radius \rstar, mass-loss rate \mdot, wind terminal speed \vinf, and the abundances of the included species. Table \ref{atomic} shows the atomic model used in this paper, which makes use of `super' levels to reduce the number of levels whose atomic populations must be solved for \citep{Anderson_1989, hill_mill}. We assume a Fe mass fraction of 1.7 $\times10^{-4}$, i.e., $\sim 0.1$ of the solar value as expected for PHL~293B. We also assume a He mass fraction of 0.5 ($\sim 1.8$ of the solar value), which is typical of LBVs \citep[e.g.][]{Groh2009a}. Since only a handful of diagnostic lines are present, both the He and Fe abundances are assumed rather than derived values. In this temperature regime, the He and Fe lines are affected by a small change in stellar and/or wind parameters \citep{boian_2018_catch}. This is specially relevant for Fe, since we assume a solar-scaled Fe/O ratio, and this may not hold depending on the chemical evolution history of the galaxy. Since the derived metallicity of PHL~293B is based on nebular O lines \citep{Izotov_2009}, for instance a change in Fe abundance by a factor of 2 would be still consistent with our models.

For simplicity our models are unclumped, and indeed they match the strength of the observed electron-scattering wings of the H lines, which are a key clumping diagnostic \citep{Hillier1991}. We degraded the CMFGEN high-resolution synthetic spectra by convolving with a Gaussian function to match the UVES spectral resolution. Because of the high wind density, we compute both a flux temperature $T_*$ at high optical depths (at Rosseland optical depth $\tau_\mathrm{Ross}$=10), as well $T_\mathrm{eff}$ at the photosphere (where $\tau_\mathrm{Ross}$=2/3). 

\begin{table}
\centering
\caption{ CMFGEN atomic model used in the analysis of PHL~293B.}
\label{atomic}
\begin{tabular}{c c c}
\hline
Species & No. of super-levels & No. of atomic levels \\
\hline
\ion{H}{i} & 20 & 30 \\
\ion{He}{i} & 40 & 45 \\
\ion{He}{ii} & 22 & 30 \\
\ion{C}{i} & 38 & 80 \\
\ion{C}{ii} & 39 & 88 \\
\ion{C}{iii} & 32 & 59 \\
\ion{N}{i} & 44 & 104 \\
\ion{N}{ii} & 157 & 442 \\
\ion{N}{iii} & 42 & 158 \\
\ion{O}{i} & 69 & 161 \\
\ion{O}{ii} & 26 & 80 \\
\ion{O}{iii} & 33 & 92 \\
\ion{Na}{i} & 18 & 44 \\
\ion{Mg}{i} & 37 & 57 \\
\ion{Mg}{ii} &18 & 45 \\
\ion{Al}{ii} & 38 & 58 \\
\ion{Al}{iii} & 17 & 45 \\
\ion{Si}{ii} & 22 & 43 \\
\ion{Si}{iii} & 20 & 34 \\
\ion{Si}{iv} & 22 & 33 \\
\ion{Ca}{i} & 23 & 39 \\
\ion{Ca}{ii} & 17 & 46 \\
\ion{Ti}{ii} & 33 & 314 \\
\ion{Ti}{iii} & 33 & 380 \\
\ion{Fe}{i} & 69 & 214 \\
\ion{Fe}{ii} & 67 & 403 \\
\ion{Fe}{iii} & 48 & 346 \\
\ion{Ni}{ii} & 29 & 204 \\
\ion{Ni}{iii} & 28 & 220 \\
\hline                                      
\end{tabular}
\end{table}

Due to its distance of 23.1~Mpc, the LBV is spatially unresolved from the underlying stellar population of PHL~293B in seeing-limited ground-based data. Therefore, we created a grid of models with varying contribution from the LBV and (flat) background galaxy to the total flux. To determine the best-fit parameters, we match simultaneously the continuum and H-$\alpha$, H-$\beta$, and \ion{Fe}{II}~5169~\AA\ lines. We also ensure that \ion{He}{I}~5876~\AA\ matches the observed level (Fig. \ref{fig: model}). A simultaneous match is needed because all the diagnostics above depend on multiple model parameters, such as the luminosity, mass-loss rate, and effective temperature. We describe below our method to simultaneously determine the stellar and wind parameters of massive stars with cool, dense winds.

For fitting the absolute level of the continuum, we use a conservative criterion and compare the mean absolute flux of our models in the region $6470-6520$~\AA\ to the observations, and only retain models that are within $\pm5\%$ of the observed value.

\begin{figure}
\begin{center} 
\includegraphics[scale=0.215]{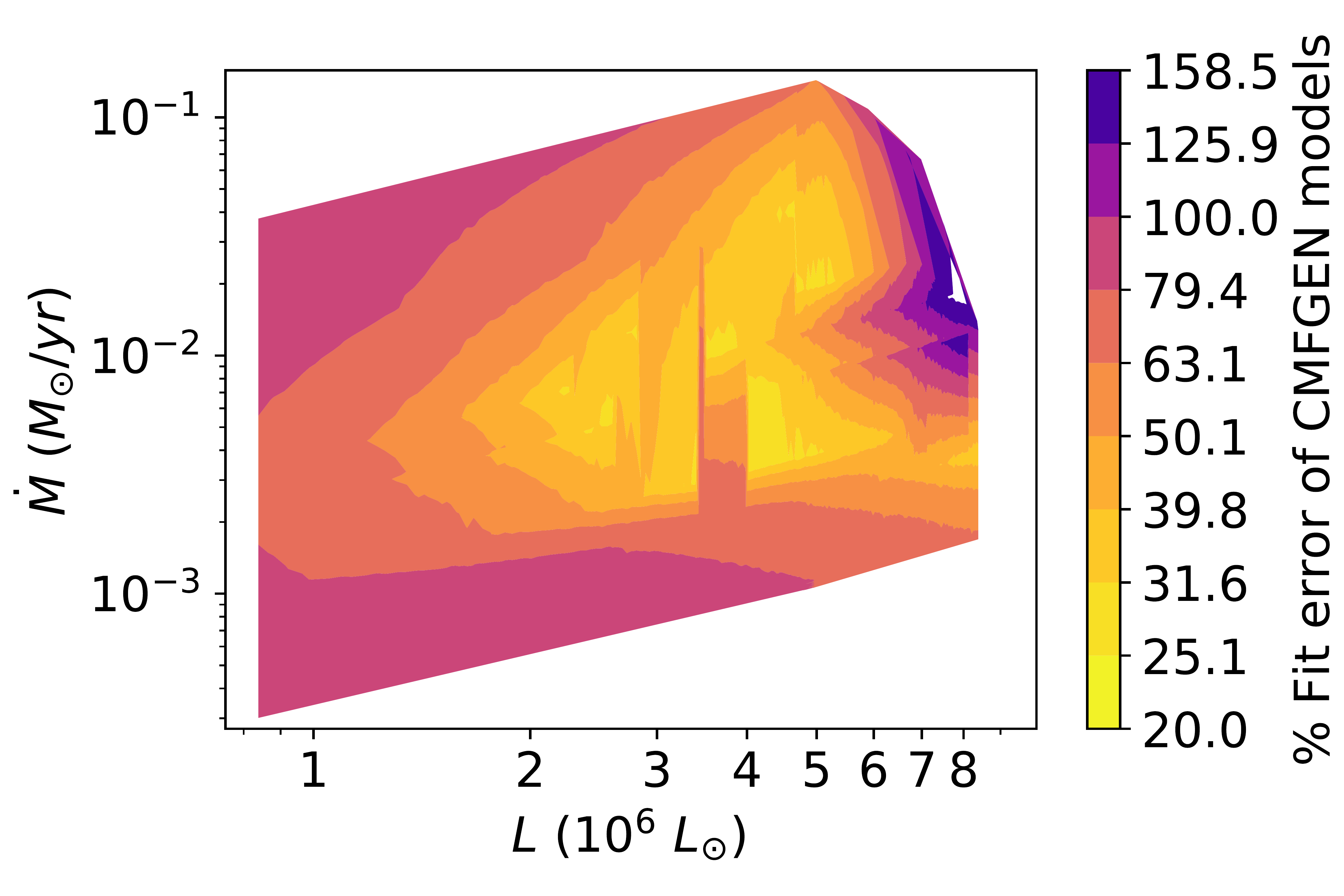}
\caption{Grid of best-fitting CMFGEN models of varying stellar mass-loss rates and luminosities. The colour denotes the fit error of the models. }
\label{fig:combo_err_grid}
\end{center}
\end{figure}

We then compare the equivalent width of the H-$\alpha$, H-$\beta$, and \ion{Fe}{II}~5169~\AA\ emission lines to the 2002 spectrum, giving equal weights to each line when computing the fit error. We chose these lines as they are the only ones where the broad component is visible with enough signal-to-noise ratio. The white regions in Figure~\ref{fig: model} show the wavelength range considered in this calculation for the H-$\alpha$, H-$\beta$ (upper panels) and \ion{Fe}{II}~5169~\AA\ lines (lower left panel). We neglect the narrow component of the Balmer lines as this is strongly affected by the background galaxy. We define our best fit models as those which produce a fit error of $<$ 25\%. These models produce a good fit to the observations, producing similar P~Cygni emission and absorption components (Figure~\ref{fig: model}). Figure \ref{fig:combo_err_grid} shows the luminosity and mass-loss rate of our CMFGEN models, with the colour indicating the percentage error for the fit. Based on the best fit models, we estimate that the LBV contributes $20-47$\% of the total flux collected in the 1\arcsec\ aperture used in the ground-based spectroscopic observations, with the rest coming from the underlying stellar population. Note that the contribution of the LBV to the total flux would be much smaller for larger apertures, e.g. those used in the ground-based photometry of \citet{Burke2020}.

\begin{figure}
\includegraphics[scale=0.46]{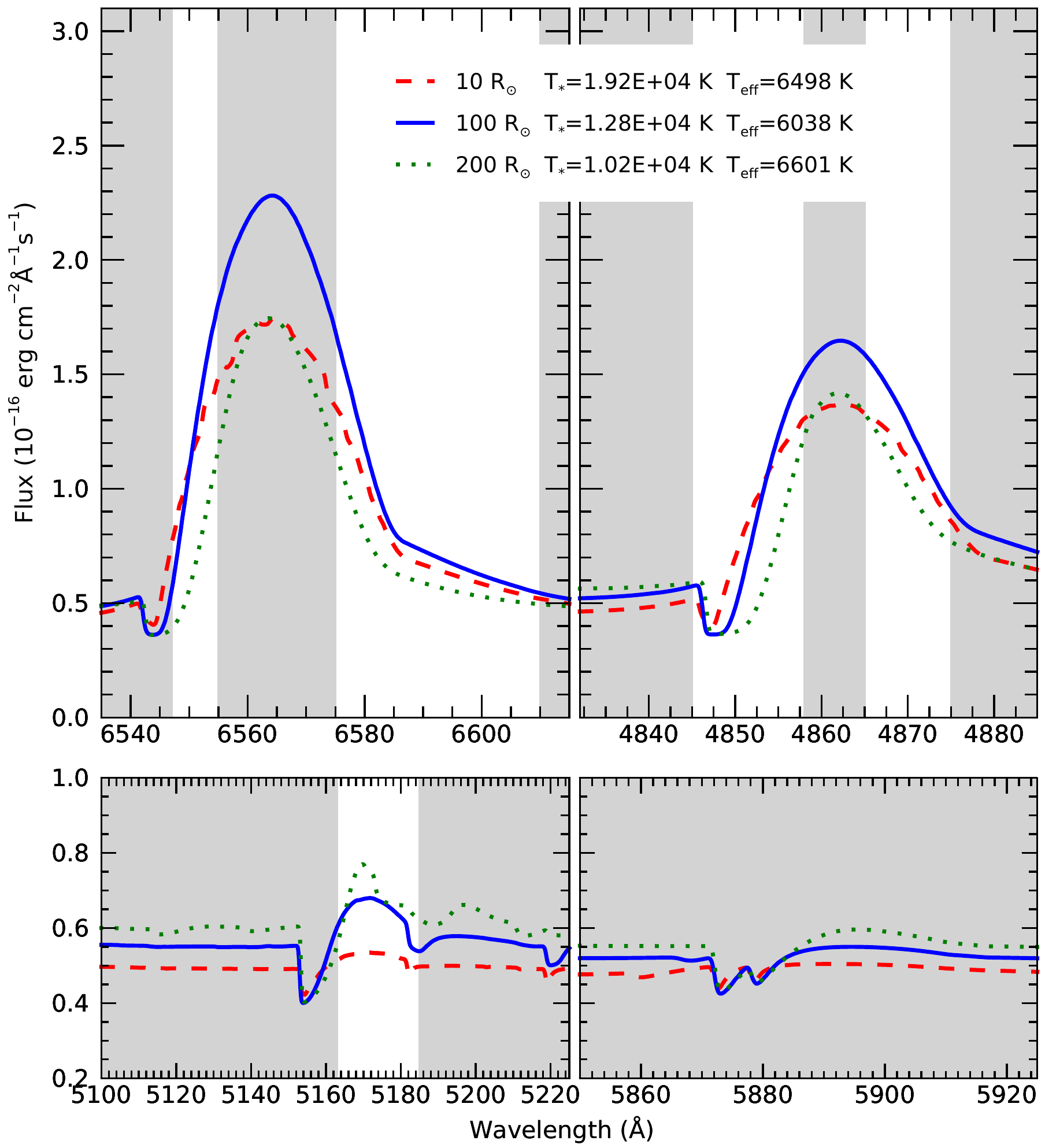}
\caption{ CMFGEN radiative transfer models for different stellar radii. The solid-blue spectrum belongs to one of our best fit models with the following parameters; \rstar=100$R_{\odot}$, $L_* = 2.52 \times 10^6 ~L_{\odot}$ and $\dot{M} = 0.008~M_{\odot}$~yr$^{-1}$. We compare this to two otherwise identical models but with $\rstar=10~\rsun$ (dashed red line) and $\rstar=200~\rsun$ (dotted green line).}  
\label{fig: RADIUS}
\end{figure}  

 \begin{figure*}
\begin{center} 
\includegraphics[scale=0.54]{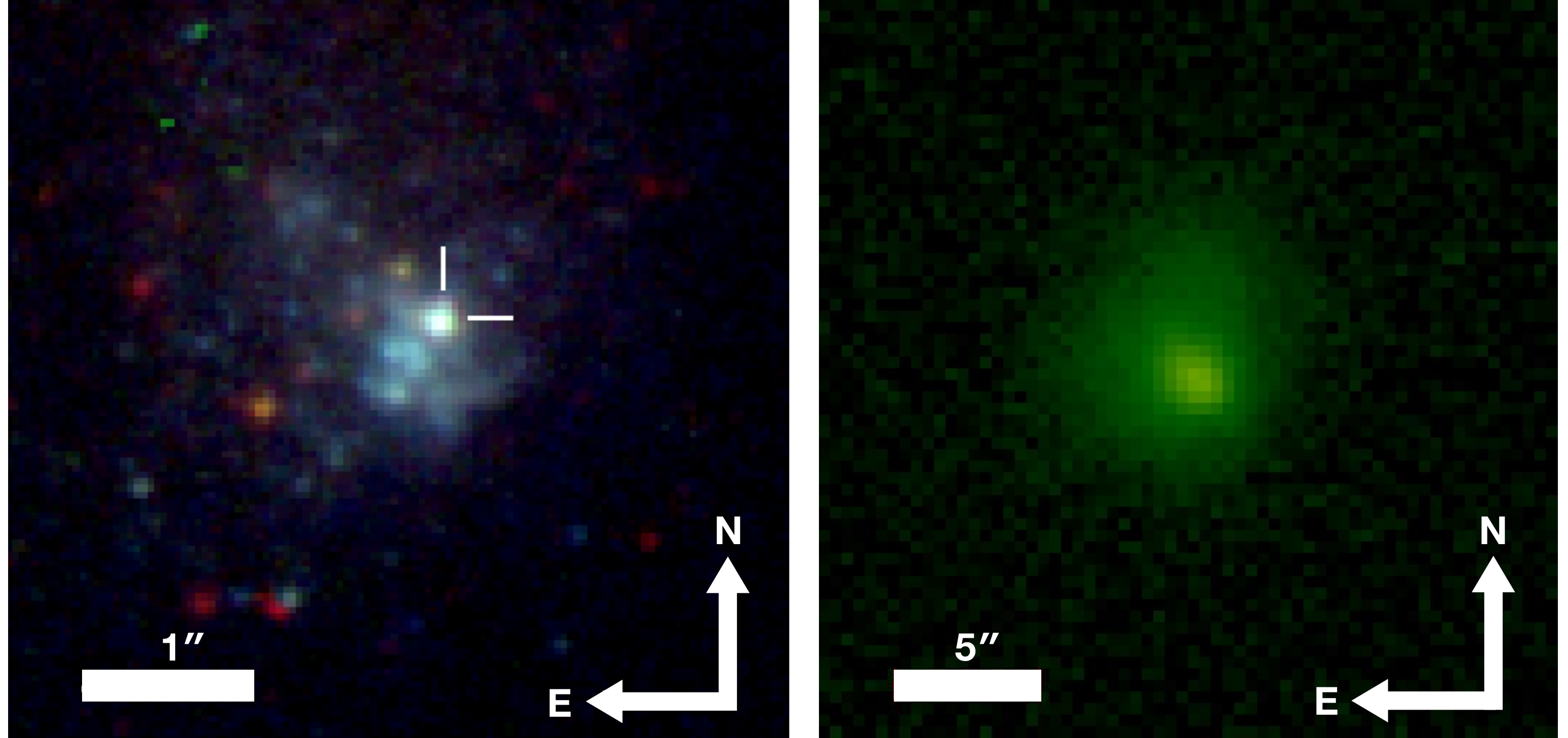}
\caption{Left: High-spatial resolution colour-composite image of PHL obtained with {\it HST} WFC3 on 2010-10-31 (GO-12018; PI Prestwich). Right: SDSS $g$-band image of PHL~293B taken on 2003-10-24. The spatial scales are indicated by the horizontal bar.}
\label{fig:PHOTO}
\end{center}
\end{figure*}  

Our best fitting CMFGEN models indicate that before its disappearance, the LBV had $L_* = 2.5-3.5 \times 10^6 ~L_{\odot}$, $\dot{M} = 0.005-0.020 ~M_{\odot}$~yr$^{-1}$, an effective temperature $T_\mathrm{eff} = 6000-6800$~K, and $T_\mathrm{*} = 9500-15000$~K. While models with $L_* = 3.5-5.0 \times 10^6 ~L_{\odot}$ are consistent with the ground-based data (Fig. \ref{fig:combo_err_grid}), they are too bright compared to the {\it HST} observations, as will be discussed. A stellar wind velocity of 1000~km~s$^{-1}$ is required by our models to reproduce the broad emission and P-Cygni absorption component of the hydrogen and iron lines. This velocity is much faster than the 40~km~s$^{-1}$ outflow observed for the extreme Red Supergiant (RSG) VY~CMa \citep{SMITH_2009_RSG} or the typical velocities of $50-300$~km~s$^{-1}$ observed in S~Doradus outbursts of LBVs \citep{vanGenderen2001}. The stellar and wind parameters strongly suggest that the LBV was in an eruptive state. 

 As the outflow of the LBV is very dense, its hydrostatic radius $R_{\star}$  is difficult to constrain. Figure \ref{fig: RADIUS} shows that over a large range of radii, the spectral morphology shows little variation. This is similar to what has been found for Eta Car, which also possesses a high-density wind \citep{Hillier_2001, Hillier_2006, GROH_2012MNRAS}. \citet{Hillier_2001} finds that the dense wind of Eta Car prevents the temperature of the underlying star from being well determined. Minor variations in line strength due to largely differing choices of hydrostatic radius and consequently temperature could instead be caused by a relatively small change (30-50\%) in the mass-loss rate or luminosity. A different He fraction could also alter emission line strengths in this manner. In the case of PHL~293B, large changes from our reference value of 100 $R_{\odot}$ would require greater mass-loss rates to fit the observations, as shown by Figure \ref{fig: RADIUS}. In this sense, our quoted values of \mdot are lower limits. We also omit the narrow region of the Balmer lines in our analysis due to its galactic origin in the observations. This further softens the importance of the chosen radius as the majority of variation resulting from changing the radius is within this component. In summary, our conclusion that the LBV possessed an extremely dense wind is not affected by the choice of \rstar.

To investigate the photometric changes that would be expected in the scenario of a disappearing star, we compute synthetic photometry of our best-fitting CMFGEN LBV models in the SDSS and {\it HST} filters. We find that our LBV models have apparent magnitudes of $m_\text{LBV,g} = 20.20 - 21.05$~mag and $m_\text{LBV,r}= 19.79 - 20.56$~mag in the SDSS $g$ and $r$ filters, respectively. In the {\it HST} F336W, F438W, F606W, and F814W filters, we obtain  $m_\text{LBV,F336W}= 19.52 - 20.20$ mag,  $m_\text{LBV,F438W}=20.27 - 21.13$ mag,  $m_\text{LBV,F606W}=19.92 - 20.72$ mag, and $m_\text{LBV,F814W}=19.79 - 20.72$~mag, respectively.

We determine the following relation for the expected change in apparent magnitude taken pre- and post-LBV disappearance,
\begin{equation}
   \Delta m=m_{\text{post}}-m_{\text{pre}} =  -2.5~ \text{log} \left( 1-10^{\frac{m_{\text{pre}}-m_{\text{LBV}}}{2.5}}   \right) ,  
  \label{eq:q}
\end{equation}
where $m_{\text{pre}}$ and $m_{\text{post}}$ are the apparent magnitudes of the galaxy in a given filter before and after the LBV disappeared, respectively, and $m_\text{LBV}$ is the apparent magnitude of the LBV.

We use the photometry from \citet{Burke2020} to estimate $m_{\text{pre}}$ for both the SDSS $g$ and $r$ bands. We interpolate the SDSS 2005 and 2013 observations and find $m_\text{pre,g} \simeq m_\text{pre,r}\simeq 17.83$~mag. Using Equation~\ref{eq:q}, we obtain $\Delta m$ of about $0.06-0.13$~mag and $0.09-0.19$~mag in the $g$ and $r$ bands, respectively, if the LBV was to completely disappear. These variations are broadly consistent with the light curve of \citet{Burke2020}. This SDSS and DES ground-based photometry uses a large aperture of $5\arcsec$ diameter and a large fraction of the underlying galaxy flux is thus included in the aperture \citep[see Fig.~1 of ][]{Burke2020}. The galactic component also captured within the aperture will have the effect of diluting the variation in the measured apparent magnitude if the LBV was to disappear. 

Figure \ref{fig:PHOTO} shows a high-spatial resolution image of PHL~293B obtained with HST (left panel) compared to an SDSS image of lower spatial resolution (right panel). Because of the high spatial resolution of the HST observations, we were able to extract photometry using a much smaller aperture than those of SDSS and DES. For that reason, the LBV contributes to a much higher fraction of the flux in the {\it HST} observations. Some of our best-fitting CMFGEN models are consistent with the magnitude and colours observed with {\it HST}, implying that the object for which we extracted 2010 {\it HST} photometry may actually be the LBV itself. Fainter LBV models are also consistent with the data. Because of the smaller aperture, if the LBV disappeared we predict much higher magnitude changes in future {\it HST} data, even for the faint models. We find lower limits for the $\Delta m$ in each filter, obtaining $\Delta m_{\text{F336W}}=1.07$, $\Delta m_{\text{F438W}}=1.54$, $\Delta m_{\text{F606W}}=1.26$, $\Delta m_{\text{F814W}}=0.81$ mags. High resolution spatial observations are therefore warranted to constrain the photometric variability of the LBV and potentially verify its complete disappearance. 

\begin{figure}
\begin{center} 
\includegraphics[scale=0.5]{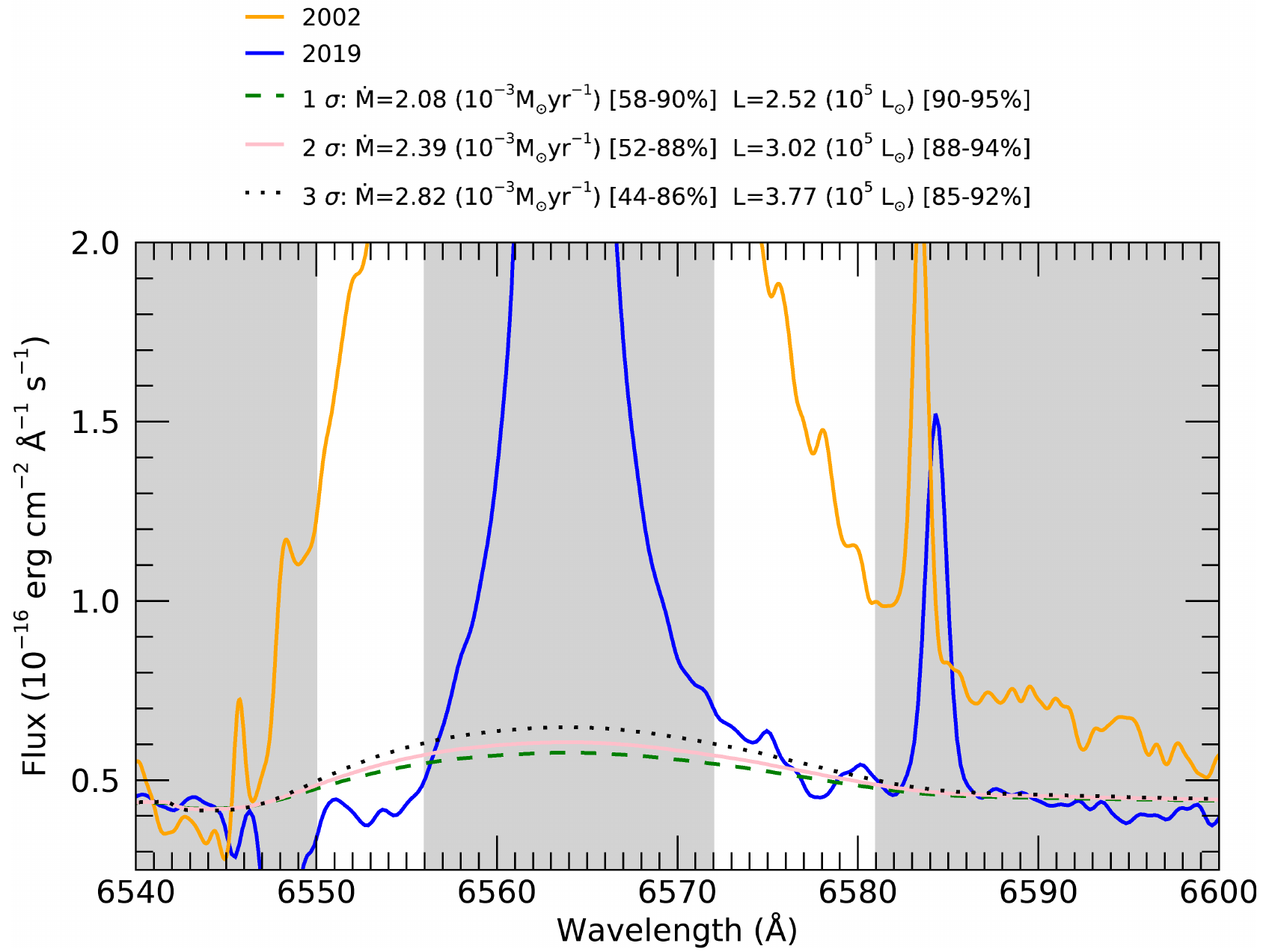}
\caption{{CMFGEN models compared to the 2019 X-shooter spectrum (blue). The models correspond to upper limits of $ L_\mathrm{surv}$ and $\dot{M}_\mathrm{surv}$ with 1 (dashed green), 2 (solid pink), 3 (dotted black) $\sigma$ confidence. The 2002 UVES spectrum (yellow) is included to highlight the significant reduction in the broad H-$\alpha$ emission. }}
\label{fig:post_diss}
\end{center}
\end{figure}

We also investigate the maximum values of the luminosity ($L_\mathrm{surv}$) and mass-loss rate ($\dot{M}_\mathrm{surv}$) that a surviving star concealed within the noise of the 2019 X-shooter spectrum could have. Assuming no dust formation and no change in temperature, we find $L_\mathrm{surv} = 3.8 \times 10^5~L_{\odot}$ and $\dot{M}_\mathrm{surv} = 2.8 \times 10^{-3}~M_{\odot}~$yr$^{-1}$ (Figure~\ref{fig:post_diss}). This would require minimum reductions in $L_*$ and $\dot{M}$ of 85--92\% and 44--86\%, respectively. However, the underlying star could be much more luminous when taking circumstellar dust into account. With the end of the eruption, the star could have become hotter and with a lower mass-loss rate. In this case, the upper limit on $L_\mathrm{surv}$ would be higher than what we quote above.

 \begin{figure}
\begin{center} 
\includegraphics[scale=0.26]{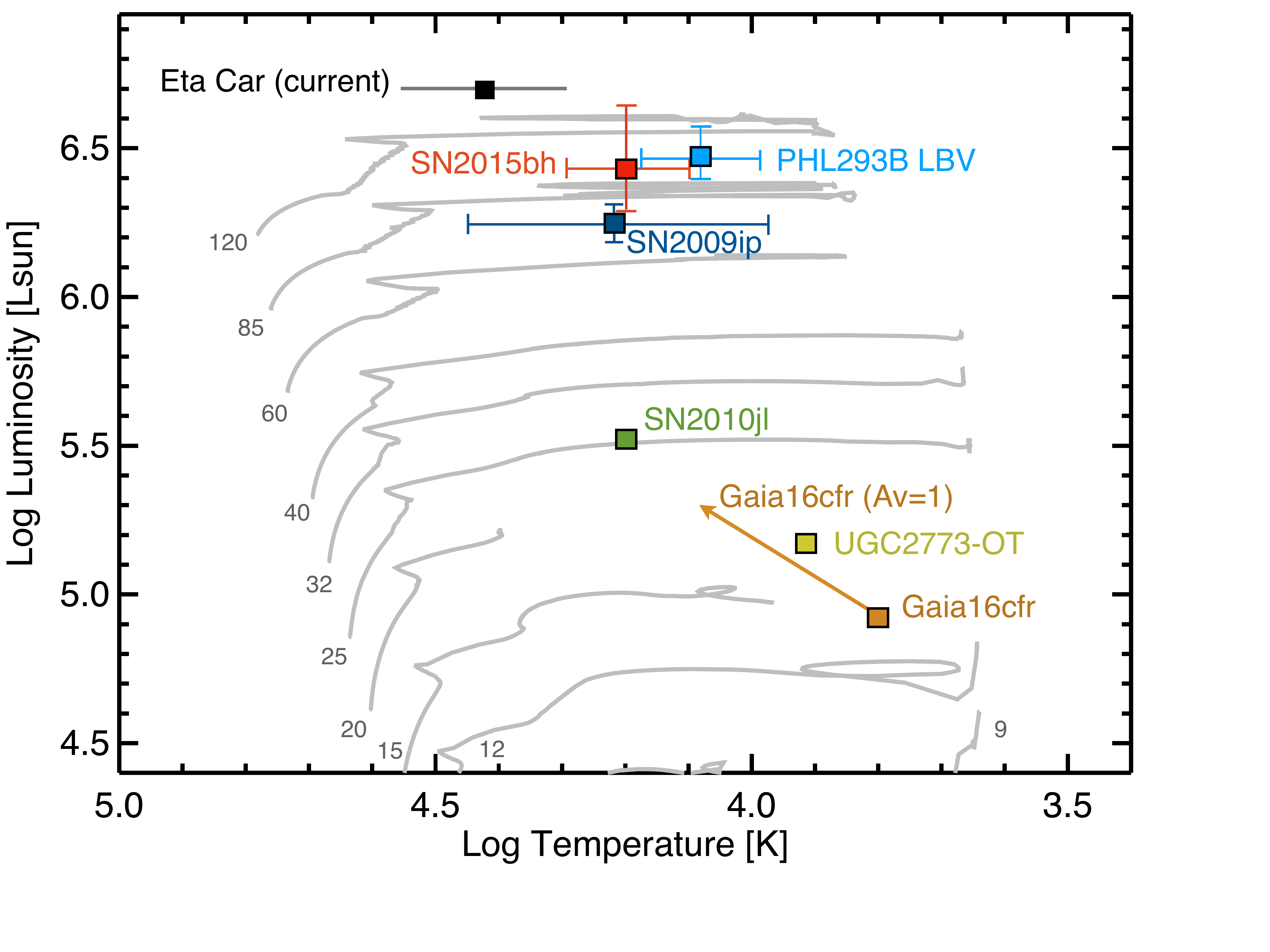}
\caption{HR diagram showing the location of the LBV in PHL~293B during 2002--2011. We also include Eta Car \citep{groh2012} and the progenitors of SN 2015bh \citep{boian_2018_catch}, SN 2009ip \citep{Smith_2010,foley11}, UGC2773-OT \citep{Smith_2010}, SN 2010jl \citep{Smith_2011}, and Gaia16cfr \citep{2018MNRAS.473.4805K}. We also show in grey the evolutionary tracks for rotating stars with initial masses in the range $9-120~M_{\odot}$ at $Z=0.0004$ \citep{Groh_GRID_Z_0.0004}. }
\label{fig:EVOL}
\end{center}
\end{figure}

\section{The fate of the massive star in PHL~293B} \label{Section_Discussion}
Based on our observations and models, we suggest that PHL~293B hosted an LBV with an eruption that ended sometime after 2011. This could have been followed by 1) a surviving star or 2) a collapse of the LBV to a BH without the production of a bright SN, but possibly with a weak transient.

One possibility is that we are seeing the end of an LBV eruption of a surviving star, with a mild drop in luminosity, a shift to higher effective temperatures, and some dust obscuration. The lack of variation of broad emission in the H-$\alpha$ and H-$\beta$ in all spectra from 2001 to 2011 would require such an eruption to have persisted for a minimum of 8.5 years. Considering the high mass-loss rate and relatively low temperatures for the outer wind from our model predictions, the dust obscuration scenario does also not necessarily require a sudden end of the eruption between the 2011 and 2016 observations. A combination of a slightly reduced luminosity and a thick dusty shell could result in the star being obscured. While the lack of variability between the 2009 and 2019 near-infrared continuum from our X-shooter spectra eliminates the possibility of formation of hot dust ($\gtrapprox 1500$~K), mid-infrared observations are necessary to rule out a slowly expanding cooler dust shell. 
 
\citet{smith_owocki2006} show that optically thick, continuum-driven outbursts could play a greater role in the mass loss of massive stars than steady, line-driven winds. Importantly, they suggest that such outbursts should be largely metallicity insensitive in comparison to line-driven winds. Our potential finding of such an outburst at a low metallicity could help confirm their hypothesis, implying that mass loss at low metallicities could be dominated by continuum-driven winds/eruptions. This conclusion would carry important implications for the final masses and SNe produced by the first stars of the Universe.

There is considerable debate on the end stages and fate of the most massive stars. Instead of surviving the eruption, the LBV in PHL~293B could have instead collapsed to a black hole (BH), with perhaps the LBV eruption signalling the end of the stellar life. Assuming that a BH has been formed, we utilise low initial metallicity (Z=0.002 and Z=0.0004) stellar evolutionary models \citep{Georgy_Z_002,Groh_GRID_Z_0.0004} to estimate the BH mass. We find that an initial mass between 85--120~$M_{\odot}$ best suits our determined parameters for the LBV. Based on these initial masses, a BH could have a mass between $40$ and $90~M_{\odot}$ through fallback, assuming no mass loss at that stage. The final BH mass depends on the rotation of the progenitor. Fast-rotating models within this initial mass range may produce a pair-instability SN rather than a core collapse to a BH. The non-detection of such a bright event, however, suggests that this was likely not the case for the LBV in PHL~293B. Evolutionary models predict that the lifetime of a star with initial mass of $120~M_{\odot}$ is between $2.88-3.90$~Myr, depending on the metallicity and rotation \citep{Georgy_Z_002,Groh_GRID_Z_0.0004}. The initial mass of the star could, however, have been significantly lower if the star was a mass gainer in a binary system. This would result in a drastically longer lifetime and lower BH mass. Determining the age of the stellar population surrounding the LBV could possibly discriminate between single and binary star evolutionary scenarios, see \citet{Smith2015}. 

A spectroscopic observation of a star immediately preceding the collapse to a BH without a bright SN would be unprecedented. LBVs span a wide range of luminosities \citep{SMITH_GAIA} and it would not be impossible for a low-luminosity, dust-reddened LBV to show eruptive behavior and perhaps collapse to a BH. This could have important consequences for N6946-BH1, for which a RSG progenitor was originally suggested \citep{Adams_2017}.

An alternative explanation for the disappearance for the LBV in PHL~293B is an undetected SN explosion. \citet{Burke2020} favour this hypothesis, suggesting that an SN IIn event occurred between 1995 and 1998, during which no photometry is available. In this case, the broad components seen in the Balmer lines between 2001--2011 would come from interaction between the SN ejecta and a dense circumstellar medium. This scenario requires that a potentially prolonged SN interaction went undetected at early times, and our data cannot rule this out.  

Both pre- and post-explosion spectra have been reported for only one SN \citep[SN~1987a;][]{Walborn_1989} and two SN candidates \citep[SN~2009ip, e.g.][]{Smith_2010,Smith_2014}; and \citep[SN~2015bh, e.g.][]{Thone_2017,boian_2018_catch}. Analysis of the pre-explosion spectra has revealed that the progenitor of SN~2015bh was an LBV star \citep{boian_2018_catch}. Interestingly, the spectra and consequently the predicted parameters of its progenitor do not greatly differ from that of the LBV in PHL~293B.

Our best-fit models for the 2001--2011 spectra place the LBV in PHL~293B at the higher $L_*$ end of the HR diagram (Figure~\ref{fig:EVOL}), in proximity to very massive LBVs such as Eta Car, and in a $T_*$ range characteristic of LBVs. It should be noted that the wind properties are stronger than those of Eta Car, with $\dot{M}_*$ being 5--20 times larger for PHL~293B. It is, however, remarkably similar to the quiescent LBV progenitor of SN~2015bh, in L, T, $\dot{M}_*$ , and $v_{\infty}$, with SN~2015bh being only slightly hotter (Figure~\ref{fig:EVOL}). SN~2015bh \citep{boian_2018_catch} is a SN candidate with a rare pre-explosion spectrum. SN~2009ip is part of the same category of events, and its progenitor also resides in a similar region of the HR diagram, however, slightly dimmer than PHL~293B, and with relatively poor $T_*$ constraints. Other LBV progenitors of luminous transients, such as SN~2010jl \citep{Stoll_2011} and Gaia16cfr \citep{2018MNRAS.473.4805K}, which have been identified in pre-explosion photometry, show much lower luminosities but consistent temperatures. An obvious difference between all mentioned transients and the PHL~293B case is the detection of a SN explosion rather than simply the fading of the star.

The case of PHL~293B is unique in the sense that several spectra were obtained shortly before its disappearance, which show spectral features that are consistent with stellar properties of an LBV in eruption. The low metallicity ($\sim0.1$ solar) of PHL~293B further amplifies its importance. Deep high-spatial resolution imaging is needed to further discriminate between the different scenarios that have been proposed. It will be highly beneficial to search for similar events in large scale surveys such as the Zwicky Transient Factory \citep[ZTF;][]{ZTF_2019} and the Large Synoptic Survey Telescope \citep[LSST;][]{LSST_2008}. Given that the majority of such events in deep surveys will be much fainter than PHL~293B and located much farther, a detailed analysis of this object in the local Universe provides an important benchmark for understanding the late-time evolution of massive stars in low metallicity environments and their remnants.

\section*{Acknowledgements}
We thank the anonymous referee for their constructive comments. AA acknowledges funding from the Provost's PhD Project Awards at Trinity College Dublin. JHG acknowledges support from the Irish Research  Council  New Foundations Award 206086.14414 'Physics of Supernovae and  Star'. IB is supported by a Trinity College Dublin Postgraduate Award. EF acknowledges funding from IRC Project 208026, Award 15330.

Based on observations made with the NASA/ESA Hubble Space Telescope, obtained from the Data Archive at the Space Telescope Science Institute, which is operated by the Association of Universities for Research in Astronomy, Inc., under NASA contract NAS5-26555. These observations are associated with program \#12018.

Funding for SDSS-III has been provided by the Alfred P. Sloan Foundation, the Participating Institutions, the National Science Foundation, and the U.S. Department of Energy Office of Science. The SDSS-III web site is http://www.sdss3.org/.

SDSS-III is managed by the Astrophysical Research Consortium for the Participating Institutions of the SDSS-III Collaboration including the University of Arizona, the Brazilian Participation Group, Brookhaven National Laboratory, Carnegie Mellon University, University of Florida, the French Participation Group, the German Participation Group, Harvard University, the Instituto de Astrof\'isica de Canarias, the Michigan State/Notre Dame/JINA Participation Group, Johns Hopkins University, Lawrence Berkeley National Laboratory, Max Planck Institute for Astrophysics, Max Planck Institute for Extraterrestrial Physics, New Mexico State University, New York University, Ohio State University, Pennsylvania State University, University of Portsmouth, Princeton University, the Spanish Participation Group, University of Tokyo, University of Utah, Vanderbilt University, University of Virginia, University of Washington, and Yale University.



\bibliographystyle{mnras}
\bibliography{AA_MAIN_NEW}


\bsp	
\label{lastpage}
\end{document}